# Secured Encryption scheme based on the Ree groups

Gennady Khalimov[1][0000-0002-2054-9186], Yevgen Kotukh [2][0000-0003-4997-620X]

[1]Kharkiv National University of Radioelectronics, Kharkiv, 61166, Ukraine hennadii.khalimov@nure.ua
[2]Yevhenii Bereznyak Military Academy, Kyiv, Ukraine yevgenkotukh@gmail.com

**Abstract.** An improved design of cryptosystem based on small Ree groups is proposed. We have changed the encryption algorithm and propose to use a logarithmic signature for the entire Ree group. This approach improves security against sequential key recovery attacks. Hence, the complexity of the key recovery attack will be defined by a brute-force attack over the entire group. In this paper, we have proved that to construct secure cryptosystems with group computations over a small finite field its needed to use 3-parametric small Ree group.

## INTRODUCTION

Within a few papers of our research on post-quantum public cryptography several approaches were considered with an aim of improving the original scheme of MST3 cryptosystem which was introduced by Lempken et al. [4]. Original design was based on random covers. Suzuki 2-group is chosen for the original design. However, there are many attack vectors found in the original design. And it became strong evidence of non-readiness for PQC era.

A more secure cryptosystem design (aka eMST3) was introduced by Svaba et al. [6]. Secret homomorphic coverage was used. After that, the general approach of strong aperiodic logarithmic signatures construction for any abelian p-groups was introduced by van Trung [9].

Our research group has joined this scientific journey [10–14] to improve MST3 cryptography, considering this cryptosystem as a PQC-ready candidate. Hence, we have considered the implementations of X-parameter groups (where X is a few parameters). We have addressed an issue with computational overheads by reducing large keys. This approach has shown a significant increase of encryption and decryption efficiency. Also, we found the way to construct a more secure cryptosystem using groups with a large order. Here, we also compute the logarithmic signature outside the group center. We do it over finite fields of small dimension.

This work shows a new encryption algorithm with bonded keys. Also, we evaluate brute-force attacks on key recovery for a cryptosystem with Ree groups.

# GENETAL ENCRYPTION SCHEME BASED ON REE GROUPS

We have a finite field $F_q$ and define a small Ree group over it as follows: $q = 3^{2n+1}$ for some $n > 0$ and $p = 3^n$ as [15]

$$Ree(q) = \langle \alpha(y), \beta(y), \gamma(y), h(\theta), I^- \mid y \in F_q, \theta \in F_q^\times \rangle.$$

We represent a subgroup $U(q)$ for the $Ree(q)$ of upper triangular matrices in the form of

$$Q(q) = \langle \alpha(y), \beta(y), \gamma(y) \mid y \in F_q \rangle.$$

We also represent each element of $Q(q)$ uniquely

$$U(x, y, z) = \alpha(x)\beta(y)\gamma(z)$$

so $Q(q) = \{U(x, y, z) \mid x, y, z \in F_q\}$, and it follows that $|Q| = q^3$. Also, $Q$ is a Sylow 3-subgroup of $Ree(q)$, and we conclude by computing the following

$$U(x_1, y_1, z_1)U(x_2, y_2, z_2) = U(x_1 + x_2, y_1 + y_2 - x_1 x_2^{3p}, z_1 + z_2 - x_2 y_1 + x_1 x_2^{3p+1} - x_1^2 x_2^{3p}),$$

$$U(x, y, z)^{-1} = U(-x, -y - z^{3p+1}, -z - xy + x^{3p+2}).$$

The center $Z(Q) = \{U(0, 0, \beta) \mid \beta \in F_q\}$.

We have subgroup $Q$ for the small group $Ree(q)$ with a greater $ordQ(q) = q^3$ than the Suzuki group orders. In original scheme Suzuki groups is used and there are isomorphic to the projective linear group $PGL(3, F_q)$, where $q = 2p^2$, $p = 2^m$ and has order $q^2$.

Here, we continue to consider the way of implementing a small Ree group-based cryptosystem [13].

We have a $Q(q) = \{U(v, w, \sigma) \mid v, w, \sigma \in F_q\}$ as a large group with $q = 3^{2n+1}$, $n > 0$, $p = 3^n$.

Tame logarithmic signatures can be defined by the following way

$$w_{(1)} = [W_{1(1)}, ..., W_{s(1)}] = (w_{ij})_{(1)} = U(0, w_{ij(1)}, 0), \quad w_{(2)} = [W_{1(2)}, ..., W_{s(2)}] = (w_{ij})_{(2)} = U(0, 0, w_{ij(2)})$$

of types $(\mu_{1(t)}, ..., \mu_{h(t)})$, $i = \overline{1, h(t)}$, $j = \overline{1, \mu_{i(t)}}$, $w_{ij(t)} \in F_q$, $t = \overline{1, 2}$ for coordinates $w$ and $\sigma$.

Then, we choose two random covers

$$v_{(1)} = [V_{1(1)}, ..., V_{s(1)}] = (v_{ij})_{(1)} = U(v_{ij(1)_v}, v_{ij(1)_w}, v_{ij(1)_\sigma}), \quad v_{(2)} = [V_{1(2)}, ..., V_{s(2)}] = (v_{ij})_{(2)} = U(0, v_{ij(2)_w}, v_{ij(2)_\sigma})$$

The type of cover is the same as $w_{(t)}$, $t = \overline{1, 2}$ respectively, $v_{ij} \in Q(q)$, $v_{ij(t)_v}, v_{ij(t)_w}, v_{ij(t)_\sigma} \in F_q \setminus \{0\}$.

We generate $\rho_{0(k)}, \rho_{1(t)}, \ldots, \rho_{h(t)} \in Q \setminus Z$, $\rho_{i(t)} = U(\rho_{i(t)_a}, \rho_{i(t)_b}, \rho_{i(t)_c})$, $\rho_{i(t)_j} \in F_q / \{0\}$, $i = \overline{0, h(t)}$, $j = \overline{1,3}$, $t = 1, 2$ and $\rho_{h(1)} = \rho_{0(2)}$.

Calculating

$$\zeta_{(t)} = [\zeta_{1(t)}, \ldots, \zeta_{h(t)}] = (\zeta_{ij})_{(t)} = \rho_{(i-1)(t)}^{-1} \varphi((v_{ij})_{(t)})(w_{ij})_{(t)} \rho_{i(t)}, i = \overline{1, h(t)}, j = \overline{1, \mu_{i(t)}}, t = \overline{1,2},$$

where $\varphi$ is a homomorphic function $\varphi(U(v, w, \sigma)) = U(0, w, \sigma)$.

Hence, we have an output public key $[\varphi, (v_t, \zeta_t)]$, and a private key $[w_{(t)}, (\rho_{0(t)}, \ldots, \rho_{h(t)})]$, $t = \overline{1,2}$.

The next step is to proceed with encryption.

As input data, we have a message $x = U(0, x_2, x_3)$ and the public key $[\varphi, (v_t, \zeta_t)]$, $t = 1, 2$.

As output data, we have a ciphertext $(z_1, z_2, z_3)$ of the message $x$.

Generate random $a = (a_1, a_2)$, $a_1, a_2 \in Z_{|Z|}$ and calculate

$$z_1 = v'(a) \cdot x = v_1'(a_1) \cdot v_2'(a_2) \cdot x, \quad z_2 = \zeta'(a) = \zeta_1'(a_1) \cdot \zeta_2'(a_2), \quad z_3 = \varphi(v_2'(a_2)).$$

We have $(z_1, z_2, z_3)$ as an output.

The next step is to proceed with decryption.

As input data, we have a ciphertext $(z_1, z_2, z_3)$ and private key $[w_{(t)}, (\rho_{0(t)}, \ldots, \rho_{h(t)})]$, $t = 1, 2$.

As an output data we have the message $x$ corresponding to the ciphertext $(z_1, z_2, z_3)$.

To perform the decryption of the message $x$, we are going to restore random numbers $a = (a_1, a_2)$. We know the parameter $v_{(1)_1}(a_1)$ from the $z_1$ and it is included in the second component of $z_2$.

Let's generate $C^{(1)}(a_1, a_2) = \rho_{0(1)} \cdot z_2 \rho_{s(2)}^{-1}$, $C^*(a) = \varphi(z_1)^{-1} C^{(1)}(a_1, a_2)$.

We may restore $a_1$ using $w_{(1)}(a_1)^{-1}$. We simply remove $\zeta_1'(a_1)$ from $z_2$

$$z_2^{(1)} = \zeta_1'(a_1)^{-1} z_2 = \zeta_2'(a_2).$$

We calculate

$$C^{(2)}(a_2) = \rho_{0(2)} \cdot z_2 \rho_{s(2)}^{-1}, \quad C^*(a) = C^{(2)}(a_2) z_3^{-1} = C^{(2)}(a_2)$$

and restore $a_2$ with $w_{(2)}(a_2)$ using $w_{(2)}(a_2)^{-1}$.

We recover $a = (a_1, a_2)$ and the message $x$ from $z_1$

$$x = v'(a_1, a_2)^{-1} \cdot z_1.$$

This general approach was confirming its results in [13]. However, it has several significant drawbacks.

First, the keys $a_1$ and $a_2$ allow sequential key recovery attack. Key recovery of $a_1$ over a brute force attack based on brute force $a_1'$ can be performed based on the computing $v_1'(a_1')$ followed by value comparison $z_1$ in coordinate $v$ within a next equation

$$z_1 = v'(a') \cdot x = U\left(v_{(1)_v}(a_1'), v_{(1)_w}(a_1') + v_{(2)_w}(a_2') + x_w, *\right).$$

Also, iterating and finding $a_1'$ have no dependencies on the value $a_2$. It is possible to recover a Key $a_2$ through computing $v_2'(a_2')$ and comparison with $z_3$ in coordinate $\sigma$ within a next equation

$$z_3 = f(v_2'(a_2')) = U\left(0, 0, v_{(2)_w}(a_2')\right).$$

We evaluate the complexity of the attack on the keys $a = (a_1, a_2)$ equals to $2q$.

Also, the encryption algorithm does not use the entire volume of the group definition. The Ree group is only the deriver group $Q_1(q) = \{U(0, w, \sigma) | w, \sigma \in F_q\}$, which has $|Q_1(q)| = q^2$, what determines the size of the message when encrypted $|x| = q^2$.

## PROPOSED APPROACH

Our analysis has shown that extension of the logarithmic signature to the whole Ree group will eliminate these shortcomings. Let's have $Q(q) = \{U(v, w, \sigma) | v, w, \sigma \in F_q\}$, with $|Q| = q^3$. We changed the encryption algorithm by implementation of bonded keys of logarithmic signatures. It improves security against a sequential recovery attack.

*Let's describe the proposed scheme as follows*

Let $Q(q) = \{U(v, w, \sigma) | v, w, \sigma \in F_q\}$ be a large group with $q = 3^{2n+1}$, $n > 0$, $p = 3^n$.

Next step is to proceed with key generation.

The tame logarithmic signatures should be chosen $w_{(t)} = [W_{1(t)}, ..., W_{h(t)}] = (w_{ij})_{(t)}$, $(w_{ij})_{(t)} \in Q(q)$ of type $(\mu_{1(t)}, ..., \mu_{h(t)})$, $i = \overline{1, h(t)}$, $j = \overline{1, \mu_{i(t)}}$, $w_{ij(t)} \in F_q$, $t = \overline{1, 3}$. Group element $(w_{ij})_{(t)}$ has a value in only one coordinates $v$, $w$, or $\sigma$, respectively. For example, $(w_{ij})_{(1)} = U(w_{ij(t)_v}, 0, 0)$.

Choose a random covers $v_{(t)} = [V_{1(t)},...,V_{h(t)}] = (v_{ij})_{(t)} = U(v_{ij(t)_v}, v_{ij(t)_w}, v_{ij(t)_\sigma})$ of the same types as $w_{(t)}$, where $v_{ij} \in Q(q)$, $v_{ij(t)_v}, v_{ij(t)_w}, v_{ij(t)_\sigma} \in F_q \setminus \{0\}$, $t = \overline{1,3}$, $j = \overline{1, \mu_{i(t)}}$, $i = \overline{0, h(t)}$.

We generate $\rho_{0(t)}, \rho_{1(t)},..., \rho_{h(t)} \in Q \setminus Z$, $\rho_{i(t)} = U(\rho_{i(t)_v}, \rho_{i(t)_w}, \rho_{i(t)_\sigma})$, $\rho_{i(t)_v}, \rho_{i(t)_w}, \rho_{i(t)_\sigma} \in F_q / \{0\}$, $t = \overline{1,3}, i = \overline{0, h(t)}$.

Let's $\rho_{h(t-1)} = \rho_{0(t)}$, $t = \overline{1,3}$.

We generate the homomorphic function defined by

$$\varphi_1(U(v,w,\sigma)) = U(0,w,\sigma), \quad \varphi_2(U(v,w,\sigma)) = U(0,0,\sigma).$$

Next, we calculate

$$\varsigma_{(1)} = [\varsigma_{1(1)},...,\varsigma_{h(1)}] = (\varsigma_{ij})_{(1)} = \rho^{-1}_{(i-1)(1)} (v_{ij})_{(1)} (w_{ij})_{(1)} \rho_{i(1)}, \quad j = \overline{1, \mu_{i(1)}}, \quad i = \overline{1, h(1)},$$

$$\varsigma_{(2)} = [\varsigma_{1(2)},...,\varsigma_{h(2)}] = (\varsigma_{ij})_{(2)} = \rho^{-1}_{(i-1)(2)} \varphi_1\left((v_{ij})_{(2)}\right)(w_{ij})_{(2)} \rho_{i(2)}, \quad j = \overline{1, \mu_{i(2)}}, \quad i = \overline{1, h(2)},$$

$$\varsigma_{(3)} = [\varsigma_{1(3)},...,\varsigma_{h(3)}] = (\varsigma_{ij})_{(3)} = \rho^{-1}_{(i-1)(3)} \varphi_2\left((v_{ij})_{(3)}\right)(w_{ij})_{(3)} \rho_{i(3)}, \quad j = \overline{1, \mu_{i(3)}}, \quad i = \overline{1, h(3)},$$

where

$$v_{ij(1)} w_{ij(1)} = U(v_{ij(1)_v}, v_{ij(1)_w}, v_{ij(1)_\sigma}) U(w_{ij(1)_v}, 0, 0) = U(v_{ij(1)_v} + w_{ij(1)_v}, *, *),$$

$$\varphi_1(v_{ij(2)}) w_{ij(2)} = U(0, v_{ij(2)_w}, v_{ij(2)_\sigma}) U(0, w_{ij(2)_w}, 0) = U(0, v_{ij(2)_w} + w_{ij(2)_w}, w_{ij(2)_\sigma}),$$

$$\varphi_2(v_{ij(3)}) w_{ij(3)} = U(0, 0, v_{ij(3)_\sigma}) U(0, 0, w_{ij(3)_\sigma}) = U(0, 0, v_{ij(3)_\sigma} + w_{ij(3)_\sigma}).$$

We have an output public key $[\varphi_1, \varphi_2, (v_t, w_t)]$, and a private key $[w_{(t)}, (\rho_{0(t)},...,\rho_{h(t)})]$, $t = \overline{1,3}$.

Next step is to proceed with encryption.

Let $x = U(x_v, x_w, x_\sigma)$ be the message, $x \in Q(q)$. Choose a random $a = (a_1, a_2, a_3)$, $a_t \in Z_{|Z|}$, $t = \overline{1,3}$.

Then, we proceed with encryption

Representation $a' = \psi(a_1, a_2, a_3) = (a_1', a_2', a_3')$ is used for encryption key.

Calculating $z_1 = v'(a') \cdot x = v_1'(a_1') \cdot v_2'(a_2') \cdot v_3'(a_3') \cdot x$.

Calculating component $z_2$.

$$\varsigma(a) = \varsigma_1'(a_1) \cdot \varsigma_2'(a_2) \cdot \varsigma_3'(a_3),$$

$$\varsigma(a) = U\left(\rho_{0(1)}^{-1} + \sum_{\substack{i=1,\\j=a_{i(1)}}}^{h(1)} (v_{ij(1)_v} + w_{ij(1)_v}) + \rho_{s(3)}, \sum_{\substack{i=1,\\j=a_{i(2)}}}^{h(2)} (v_{ij(2)_w} + w_{ij(2)_w}) + *, \sum_{\substack{i=1,\\j=a_{i(3)}}}^{h(3)} (v_{ij(3)_\sigma} + w_{ij(3)_\sigma}) + *\right)$$

$$z_2 = \varsigma(a) \cdot \varphi_2(v_3'(a_3)) \cdot \varphi_1(v_3'(a_3)) \cdot \varphi_1(v_2'(a_2)),$$

where

$$\varphi_1(v_t{}'(a_t)) = \prod_{\substack{i=1,\\ j=a_{i(k)}}}^{h(t)} U(v_{ij(t)_v},0,0) = U\left(\sum_{\substack{i=1,\\ j=a_{i(t)}}}^{h(t)} v_{ij(t)_v},*,*\right),\ t=2,3$$

$$\varphi_2(v_t{}'(a_t)) = \prod_{\substack{i=1,\\ j=a_{i(t)}}}^{h(t)} U(0,v_{ij(t)_w},0) = U\left(0,\sum_{\substack{i=1,\\ j=a_{i(t)}}}^{h(t)} v_{ij(t)_w},0\right), t=3$$

and

$$z_2 = U\left(\rho_{0(1)}{}^{-1} + \sum_{t=1}^{3}\sum_{\substack{i=1,\\ j=a_{i(1)}}}^{h(1)} v_{ij(1)_v} + \sum_{\substack{i=1,\\ j=a_{i(1)}}}^{h(1)} w_{ij(1)_v} + \rho_{h(3)}, \sum_{\substack{i=1,\\ j=a_{i(2)}}}^{h(2)} (v_{ij(2)_w} + w_{ij(2)_w}) + \sum_{\substack{i=1,\\ j=a_{i(3)}}}^{h(3)} v_{ij(3)_w} + *, \sum_{\substack{i=1,\\ j=a_{i(3)}}}^{h(3)} (v_{ij(3)_\sigma} + w_{ij(3)_\sigma}) + *\right)$$

Calculating

$$\lambda(a) = v_1{}'(a_1)\cdot\varphi_1(v_2{}'(a_2))\cdot\varphi_1(v_3{}'(a_3)),$$

$$z_3 = \lambda(a)\varphi_1(v_3{}'(a_3))\cdot\varphi_1(v_2{}'(a_2)),$$

where

$$\varphi_1(v_t{}'(a_t)) = \prod_{\substack{i=1,\\ j=a_{i(t)}}}^{h(t)} U(0,v_{ij(t)_v},v_{ij(t)_\sigma}) = U\left(0,\sum_{\substack{i=1,\\ j=a_{i(t)}}}^{h(t)} v_{ij(t)_v}, \sum_{\substack{i=1,\\ j=a_{i(t)}}}^{h(t)} v_{ij(t)_\sigma}\right)$$

for $t=2,3$ and

$$z_3 = U\left(\sum_{t=1}^{3}\sum_{\substack{i=1,\\ j=a_{i(t)}}}^{h(t)} v_{ij(k)_v}, \sum_{t=2}^{3}\sum_{\substack{i=1,\\ j=a_{i(t)}}}^{h(t)} v_{ij(t)_w} + *, \sum_{t=2}^{3}\sum_{\substack{i=1,\\ j=a_{i(t)}}}^{h(t)} v_{ij(t)_\sigma} + *\right).$$

So, we have an output $(z_1,z_2,z_3)$.

Next step is to proceed with decryption.

To decrypt a message $x$ random numbers $a=(a_1,a_2,a_3)$ is needed to be restored.

We generate

$$C(a) = \rho_{0(1)}z_2 z_3{}^{-1}\rho_{s(3)}{}^{-1}$$

$$C(a_1,a_2,a_3) = \rho_{0(1)}U\left(\rho_{0(1)}{}^{-1} + \sum_{t=1}^{3}\sum_{\substack{i=1,\\ j=a_{i(1)}}}^{h(1)} v_{ij(t)_v} + \sum_{\substack{i=1,\\ j=a_{i(1)}}}^{h(1)} w_{ij(1)_v} + \rho_{h(3)},*,*\right)$$

$$U\left(\sum_{t=1}^{3}\sum_{\substack{i=1,\\ j=a_{i(t)}}}^{h(t)} v_{ij(k)_v}, \sum_{t=2}^{3}\sum_{\substack{i=1,\\ j=a_{i(t)}}}^{h(t)} v_{ij(t)_w} + *, \sum_{t=3}^{3}\sum_{\substack{i=1,\\ j=a_{i(t)}}}^{h(t)} v_{ij(t)_\sigma} + *\right)^{-1} \rho_{h(3)}{}^{-1} = U\left(\sum_{\substack{i=1,\\ j=a_{i(1)}}}^{h(1)} w_{ij(1)_v}\rho_{h(3)},*,*\right)$$

We restore $a_1$ with $w_{(1)}(a_1) = \sum_{\substack{i=1, \\ j=a_{i(1)}}}^{h(1)} w_{ij(1)_v}$ using $w_{(1)}(a_1)^{-1}$, because $w_1$ is simple. Here we need to remove $\zeta_1'(a_1)$ from $z_2$ and $v_1'(a_1)$ from $z_3$.

We calculate

$$z_2^{(1)} = \zeta_1'(a_1)^{-1} z_2 = \zeta_2'(a_2)\zeta_3'(a_3) \cdot \varphi_2(v_3'(a_3)) \cdot \varphi_1(v_3'(a_3)) \cdot \varphi_1(v_2'(a_2)) =$$

$$U\left(\rho_{0(2)}^{-1} + \sum_{t=2}^{3}\sum_{\substack{i=1, \\ j=a_{i(1)}}}^{h(1)} v_{ij(t)_a} + \rho_{h(3)}, \sum_{\substack{i=1, \\ j=a_{i(2)}}}^{h(2)}(v_{ij(2)_b} + w_{ij(2)_b}) + \sum_{\substack{i=1, \\ j=a_{i(3)}}}^{h(3)} v_{ij(3)_b} + *, \sum_{\substack{i=1, \\ j=a_{i(3)}}}^{h(3)}(v_{ij(3)_c} + w_{ij(3)_c}) + *\right).$$

and

$$z_3^{(1)} = v_1'(a_1)^{-1} z_3 = \varphi_1(v_2'(a_2)) \cdot \varphi_1(v_3'(a_3))\varphi_1(v_3'(a_3)) \cdot \varphi_1(v_2'(a_2)) = U\left(\sum_{t=2}^{3}\sum_{\substack{i=1, \\ j=a_{i(t)}}}^{h(t)} v_{ij(t)_v}, \sum_{t=2}^{3}\sum_{\substack{i=1, \\ j=a_{i(t)}}}^{h(t)} v_{ij(t)_w} + *, \sum_{t=2}^{3}\sum_{\substack{i=1, \\ j=a_{i(t)}}}^{h(t)} v_{ij(t)_\sigma} + *\right)$$

Same calculations to be performed for $C(a_2, a_3)$

$$C(a_2, a_3) = \rho_{0(2)} z_2^{(1)}\left(z_3^{(1)}\right)^{-1}\rho_{h(3)}^{-1} = U\left(0, \sum_{\substack{i=1, \\ j=a_{i(2)}}}^{h(2)} w_{ij(2)_w}, \sum_{\substack{i=1, \\ j=a_{i(3)}}}^{h(3)}(v_{ij(3)_\sigma} + w_{ij(3)_\sigma}) + *\right)$$

We restore $a_2$ with $w_{(2)}(a_2) = \sum_{\substack{i=1, \\ j=a_{i(1)}}}^{h(2)} w_{ij(1)_v}$ using $w_{(2)}(a_2)^{-1}$, because $w_2$ is simple.

We remove the component $\zeta_2'(a_2)$ from $z_2^{(1)}$ and $\varphi_1(v_2'(a_2))$ from $z_3^{(1)}$.

$$z_2^{(2)} = \zeta_2'(a_2)^{-1} z_2^{(1)} = U\left(\rho_{0(3)}^{-1} + \sum_{t=2}^{3}\sum_{\substack{i=1, \\ j=a_{i(1)}}}^{h(1)} v_{ij(k)_v} + \rho_{h(3)}, \sum_{\substack{i=1, \\ j=a_{i(3)}}}^{h(3)} v_{ij(3)_w} + *, \sum_{\substack{i=1, \\ j=a_{i(3)}}}^{h(3)}(v_{ij(3)_\sigma} + w_{ij(3)_\sigma}) + *\right)$$

and

$$z_3^{(2)} = \varphi_1(v_2'(a_2))^{-1} z_3^{(1)} = U\left(\sum_{t=2}^{3}\sum_{\substack{i=1, \\ j=a_{i(1)}}}^{h(1)} v_{ij(k)_v}, \sum_{\substack{i=1, \\ j=a_{i(3)}}}^{h(3)} v_{ij(3)_w} + *, \sum_{\substack{i=1, \\ j=a_{i(3)}}}^{h(3)}(v_{ij(3)_\sigma} + w_{ij(3)_\sigma}) + *\right)$$

Calculating

$$C(a_3) = \rho_{0(3)} z_2^{(2)}\left(z_3^{(2)}\right)^{-1}\rho_{h(3)}^{-1} = U\left(0, 0, \sum_{i=1, j=a_{i(3)}}^{h(3)} w_{ij(3)_c}\right)$$

Restore $a_3$ with $w_{(3)}(a_3)$ using $w_{(3)}(a_3)^{-1}$.

We obtain $a' = \psi(a_1, a_2, a_3) = (a_1', a_2', a_3')$ and recover the message $x$ from $z_1$

$$x = v'(v_1', v_2', v_3')^{-1} \cdot z_1.$$

# SECURITY ANALYSIS

We introduced our results of short security analysis mainly focused on brute-force key recovery and algorithm attacks.

Our first consideration is a brute force attack on the ciphertext $z_1$. We choose $a = (a_1, a_2, a_3)$ try to decipher the text $z_1' = v'(a') \cdot x = v_1'(a_1') \cdot v_2'(a_2') \cdot v_3'(a_3') \cdot x$. The covers $v_{(t)} = [V_{1(t)}, ..., V_{s(t)}] = (v_{ij})_{(t)} = U(v_{ij(t)_v}, v_{ij(t)_w}, v_{ij(t)_\sigma})$ selected a random and value $v'(a')$ is defined by multiplication in a group without any constraints for the coordinate. The resulting vector $v'(a')$ depends on all components $v_1'(a_1'), v_2'(a_2'), v_3'(a_3')$. Iterating over key values $a = (a_1, a_2, a_3)$ has a difficulty rating $q^3$. For a practical message, attack $m$ is also unknown and has uncertainty for choice $q^3$. This makes a brute-force attack on the key not having a correct solution. If we take an attack model with a known text, then the complexity of the attack still remains equal to $q^3$.

Or second consideration is a brute force attack on the ciphertext $z_2$. Select $a = (a_1, a_2, a_3)$ to match $z_2 = \zeta(a) \cdot \varphi_2(v_3'(a_3)) \cdot \varphi_1(v_3'(a_3)) \cdot \varphi_1(v_2'(a_2))$.

Let's represent $z_2$ over components $v_i'(a_i)$

$$z_2 = U\left(\rho_{0(1)}^{-1} + \sum_{t=1}^{3} \sum_{\substack{i=1, \\ j=a_{i(1)}}}^{h(1)} v_{ij(k)_v} + \sum_{\substack{i=1, \\ j=a_{i(1)}}}^{h(1)} w_{ij(1)_v} + \rho_{s(3)}, \sum_{\substack{i=1, \\ j=a_{i(2)}}}^{h(2)} \left(v_{ij(2)_w} + w_{ij(2)_w}\right) + \sum_{\substack{i=1, \\ j=a_{i(3)}}}^{h(3)} v_{ij(3)_w} + *, *\right)$$

We highlight $(*)$ components which are defined by cross-calculations in the group operation. Group operation of the product of $\rho_{0(t)}, ..., \rho_{h(t)}$ and product of $v_{(1)_v}(a_1), w_{(1)}(a_1)$ for coordianates $v$ and product of $v_{(1)_v}(a_1), w_{(1)}(a_1), v_{(2)_w}(a_2), w_{(2)}(a_2)$ for coordinate $\sigma$ is used.

Coordinates' values are defined by calculations over vectors $v_1'(a_1), v_2'(a_2), v_3'(a_3)$. The keys $a_1, a_2, a_3$ are related, a change in any of them leads to a change $z_2$. A brute-force attack on a key $a = (a_1, a_2, a_3)$ has a complexity equal to $q^3$.

Our third consideration is a brute force attack on the ciphertext $z_3$.

We choose $a = (a_1, a_2, a_3)$ to match $z_3 = \lambda(a)\varphi_1(v_3'(a_3)) \cdot \varphi_1(v_2'(a_2))$. Let's represent $z_3$ over components $v_i'(a_i)$. We will get

$$z_3 = U\left(\sum_{t=1}^{3}\sum_{\substack{i=1,\\j=a_{i(t)}}}^{h(t)} v_{ij(t)_v}, \sum_{t=2}^{3}\sum_{\substack{i=1,\\j=a_{i(t)}}}^{h(t)} v_{ij(t)_w} + *, \sum_{t=2}^{3}\sum_{\substack{i=1,\\j=a_{i(t)}}}^{h(t)} v_{ij(t)_\sigma} + *\right)$$

We highlight $(*)$ components which are defined by cross-calculations in the group operation. Group operation with product of $v_{(1)_v}(a_1), w_{(1)}(a_1)$ coordinates $w$ and a product of $v_{(1)_v}(a_1), v_{(2)_w}(a_2)$ for coordinate $\sigma$.

Values of coordinates $z_3$ are defined by calculations over vectors $v_1'(a_1), v_2'(a_2), v_3'(a_3)$. The keys $a_1, a_2, a_3$ are also binding, a change in any of them leads to a change $z_3$. A brute-force attack on a key $a = (a_1, a_2, a_3)$ also has a complexity equal to $q^3$.

Our fourth consideration is a brute force attack on the vectors $(\rho_{0(l)}, ..., \rho_{h(l)})$. A brute-force attack on $(\rho_{0(l)}, ..., \rho_{h(l)})$ is common for MST cryptosystems and for calculations in the field $F_q$ over a group center $Z(G)$ has an optimistic lower complexity bound equalt to $q$. For our encryption algorithm, calculations are performed on the entire group $|G| = q^3$ and the complexity of a brute force attack on $(\rho_{0(l)}, ..., \rho_{h(l)})$ will be equal to $q^3$.

Our fifth consideration is an attack on the algorithm. There are many details related to the vulnerabilities of group operation or logarithmic signature itself that corresponds to this attack. Our estimation is valid for the implementation of the MST on any noncommutative group. However, complexity requires a separate analysis. As we shown in our previous papers this attack has a lot of details that come from the design of logarithmic signature and group operation.

## CONCLUSION

Within the results of this research we see that for encryption on the entire group $Q(q) = \{U(v, w, \sigma) | v, w, \sigma \in F_q\}$ with bind keys $a = (a_1, a_2, a_3)$ the small Ree groups have shown a complexity of the brute-force attack equal to $q^3$. Our proposal includes an extension of logarithmic signature to the entire Ree group $Q(q) = \{U(v, w, \sigma) | v, w, \sigma \in F_q\}$, with $|Q| = q^3$. We redesign our encryption algorithm in such a way as to bind the keys of logarithmic signatures and improve the security against a sequential recovery attack.